# Pennants for Descriptors

Howard D. White
College of Information Science and Technology
Drexel University
Philadelphia, PA 19104 USA

Philipp Mayr
GESIS, Leibniz Institute for the Social Sciences
Unter Sachsenhausen 6-8
50667 Cologne, Germany

We present a new technique for displaying the descriptors related to a descriptor across literatures, rather in a thesaurus. It has definite implications for online searching and browsing.

The technique is pennant diagrams, introduced with several kinds of bibliographic data in White (2007a, b, 2009, 2010). White's examples to date, including the one in this paper, have required data-gathering in the Dialog search system (with its Rank command), to which many information scientists lack access. However, simple pennant diagrams, using 25 co-cited author names, have also been programmed into Drexel University's AuthorWeb system (White et al. 2001; Lin et al. 2003). As designers experiment with new ways of visualizing knowledge organization systems for their users, there seems little impediment to including pennant diagrams. The only data they require are (a) a user-supplied seed term to initiate the display, (b) non-zero co-occurrence counts of every term with the seed, and (c) total frequency counts for each of these co-occurring terms in the database.

The counts in (b) and (c) are the basic input to well-known tf*idf term-weighting schemes in document retrieval. In making a pennant, the weighting scheme in Manning and Schütze (2000) is used. The counts in (b) are converted to a tf (term frequency) weight as *log(count) + 1*, and the counts in (c) are converted to an idf (inverse document frequency) weight as *log(N/count)*, where N is the total number of documents in the database (estimated if not known).

Pennants, named for the flag they resemble, are a form of algorithmic prediction. Their cognitive base is in relevance theory (RT) from linguistic pragmatics (Sperber & Wilson 1995). In RT, the relevance of a message in a particular context depends on two factors that operate simultaneously. The first is the message's *cognitive effects* on a hearer or reader: the greater the cognitive effects it produces, the greater its relevance. The second is the *processing effort* the message costs the hearer or reader: the easier it is to process, the greater its relevance. These two factors underlie the positioning of terms on the pennant. The seed term is always at the tip, and the other terms are placed (as a scatterplot) on two logarithmic axes with respect to the seed. The horizontal axis represents cognitive effects (from low at left to high at right). Placement on it represents a term's co-occurrence count with the seed: the higher the logged count, the more the term is pulled toward the seed on that axis. This predicts that the user will experience greater cognitive effects the closer a term is to the seed.

The vertical axis represents the predicted ease of processing a term (from low at bottom to high at top). Placement on it represents a term's total count in the database. The *lower* the count (before the logged weighting), the easier the term is to process in association with the seed. Why? Because terms with counts lower than the seed's tend to be *very specifically related* to the seed and hence are easy to interpret. They will often be from the same semantic hierarchy in a thesaurus. Conversely, terms with counts considerably higher than the seed's (before the logged weighting), tend to have much broader implications. They are not specifically related to the seed and, in a thesaurus, would usually come from wholly other hierarchies. In between on the vertical axis are terms that are very roughly equal in specificity to the seed, neither narrow and highly focused, nor very broad and general.

All of this is clearer in the example, which involves descriptors from H. W. Wilson's Social Sciences Abstracts on the Dialog system. The seed term at the tip is "Immigration and Emigration," and the descriptors on display co-occur with it at least 50 times—an arbitrary threshold set for this paper. The degrees of specificity of co-occurring terms are indicated as sectors A, B, and C. (These sectors are drawn manually, although it might be possible to do them algorithmically in the future.)

Many of the sector A terms derive from "migration," the same semantic root as the seed. Others link migrants to labor markets and legal issues, always salient concerns in this field of study. The sector A terms seem typical "see also" references, of the sort that nonspecialized indexers such as librarians can make because they are *easy to see*. In the language of RT, they are cognitively easy to process.

Contrast them with sector C terms, all of which might be useful in a literature search, but none of which imply "Immigration and Emigration." Here we see names of countries, broad sorts of "aspects" and "conditions," and highly general categories, such as "Women," "Family," "Youth" and "Aged." These are harder to process cognitively in relation to the seed because they are so vague and variable in their implications. In sector B the terms are also relatively broad, but many of them could fit under the even broader categories of sector C terms.

The pennant is not predicting the topics that are most relevant to the user; it is showing the user, in a single display, *the structure of existing literatures in this database*. It shows that, as literatures have grown, certain descriptors have seemed particularly relevant to indexers *in the context of this seed term*—e.g., "Government Policy," "Migration, Internal," "Immigration and Emigration Law," and "History." ("United States," a special case circled at the bottom of the map, is extremely relevant to the seed, but in this database so many immigration studies deal with the U.S. (61%) that much additional effort is needed to discover what they are actually about.) At the same time, the pennant shows almost 100 descriptors, any one of which will produce at least 50 documents if ANDed with the seed, because indexers have already made the connections with literary warrant. This is very different from looking up "Immigration and Emigration" in a thesaurus and noting other possible leads. The pennant thus acts as a recommender system for topical areas that will be fruitful to browse or search.

Pennant diagram of descriptors related to the descriptor "Immigration and Emigration" in H. W. Wilson's Social Sciences Abstracts (Dialog file 142), July 2013.